\documentclass[prl,twocolumn,aps,showpacs]{revtex4}
\usepackage{bm}
\usepackage{amsmath}
\usepackage{amssymb}
\usepackage[dvips]{color}
\usepackage[dvips]{graphics}
\usepackage[dvips]{graphicx}
\usepackage{eepic}
\usepackage{epic}
\usepackage{hangcaption}
\usepackage{footnpag}
\usepackage{pstricks}%
\usepackage{pst-node}%
\usepackage{pst-poly}
\usepackage{graphicx}
\usepackage{graphics}
\usepackage{eepic}
\usepackage{hangcaption}

\renewcommand{\i}{{\rm i}}

\renewcommand{\Im}{{\rm Im}\;}
\newcommand{\sgn}{{\rm sgn}}

\begin{document}
\title{
Layer-resolved conductivities in multilayer graphenes}
\author{Takeo Wakutsu,$^1$ Masaaki Nakamura,$^1$ and Bal\'azs D\'ora$^2$}
\affiliation{
$^1$Department of Physics, Tokyo Institute of Technology, Oh-Okayama,
Meguro-ku, Tokyo 152-8551, Japan \\
$^2$Department of Physics, Budapest University of Technology and
Economics, Budafoki \'{u}t 8, 1111 Budapest, Hungary}
%
\begin{abstract}
We study interlayer transport of multilayer graphenes in magnetic
field with various stacking structures (AB, ABC, and AA types) by
calculating the Hall and longitudinal conductivities as functions of Fermi
energy.  Their behavior depends strongly on the stacking
structures and selection of the layers.  The
Hall conductivity between different layers is no longer quantized.
Moreover, for AB stacking, the interlayer conductivity vanishes around zero energy with increasing layer separation, and shows negative values
in particular cases. The fact that longitudinal interlayer conductivity
suppressed by the magnetic field indicates that this system can be
applied as a switching device.
\end{abstract}

\maketitle

---{\it Introduction}--- Graphene has attracted increasing attention
 since its first isolation from graphite in 2004
 \cite{Novoselov-mono}. Graphene consists of two-dimensional hexagonal
 lattice of carbon atoms, whose quasiparticles are governed by a
 massless Dirac equation. A variety of unusual phenomena are observed in
 this system such as universal value of the minimum conductivity,
 anomalous quantum Hall effect, and so on
 \cite{Novoselov-QHE,Zheng,Gusynin-Hall}. In addition, bilayer graphene
 has also been studied intensely
 \cite{McCann,Novoselov-bi,Nakamura-bi,Gusynin-longi}, which is
 characterized by intrinsic Landau level degeneracy at zero energy and a
 gate tunable band gap.


Due to the qualitative differences between mono- and bilayer graphene, a
 lot of interest has been focused on multilayer graphenes to determine,
 how additional layers influence their physical response.  One of the
 most intriguing property of these systems is the variety of stacking
 structures such as AB (Bernal), ABC (rhombohedral) and AA (simple
 hexagonal) types. Graphene is usually produced by micro-mechanical
 cleavage of graphite, so that the stacking structure is usually of AB
 type, since the natural graphite falls into this category. However,
 production of graphene with other stacking types is also possible by
 recent technology such as epitaxial methods \cite{Ohta,Norimatsu}. In
 addition, AA stacking can be realized by folding of a graphene sheet
 \cite{Liu}. In terms of band structure, multilayer graphenes with more
 than 10 layers are regarded as bulk graphite \cite{Partoens}, so that
 few layer graphenes have been considered to be important systems,
 interpolating between graphene and graphite.  So far, diamagnetism
 \cite{Koshino-mag}, transport properties \cite{Nakamura-multi,
 Maassen,Yuan,L.Zhang,Nakamura-AA,Kumar,Min,MacDonald}, and energy
 spectra \cite{Koshino-ABC,Koshino-LL} have been studied for these
 systems.

\begin{figure}[t]
\begin{picture}(110,110)
\put(-10,-55){\scalebox{3}[3]{\includegraphics[width=3.0pc]{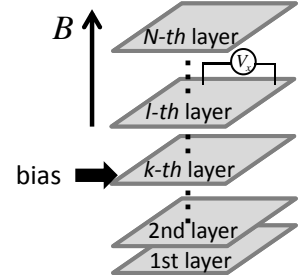}}}
\end{picture}
\caption{Schematic illustration of the layer-resolved magneto transport
between $k$-th and $l$-th layers in a multilayer graphene.}
\label{fig1}
\end{figure}

In this Letter, we investigate the transport properties between two
different layers of multilayer graphenes in magnetic field as
illustrated in Fig.~\ref{fig1}, using the Kubo formula.  Here, while
electric current is run through a given layer, the resulting voltage
drop or induced current is measured in another layer.  In order to
calculate these ``layer-resolved conductivities'', we establish the
formalisms to obtain eigenvalues and eigenstates of multilayer systems,
using the block diagonalization technique. We find many interesting
properties such as negative response and switching effect by magnetic
field, absent in monolayer graphene.

\begin{figure}[t]
\begin{picture}(100,100)
\put(-85,-20){\rotatebox{90}{\includegraphics[width=16pc]{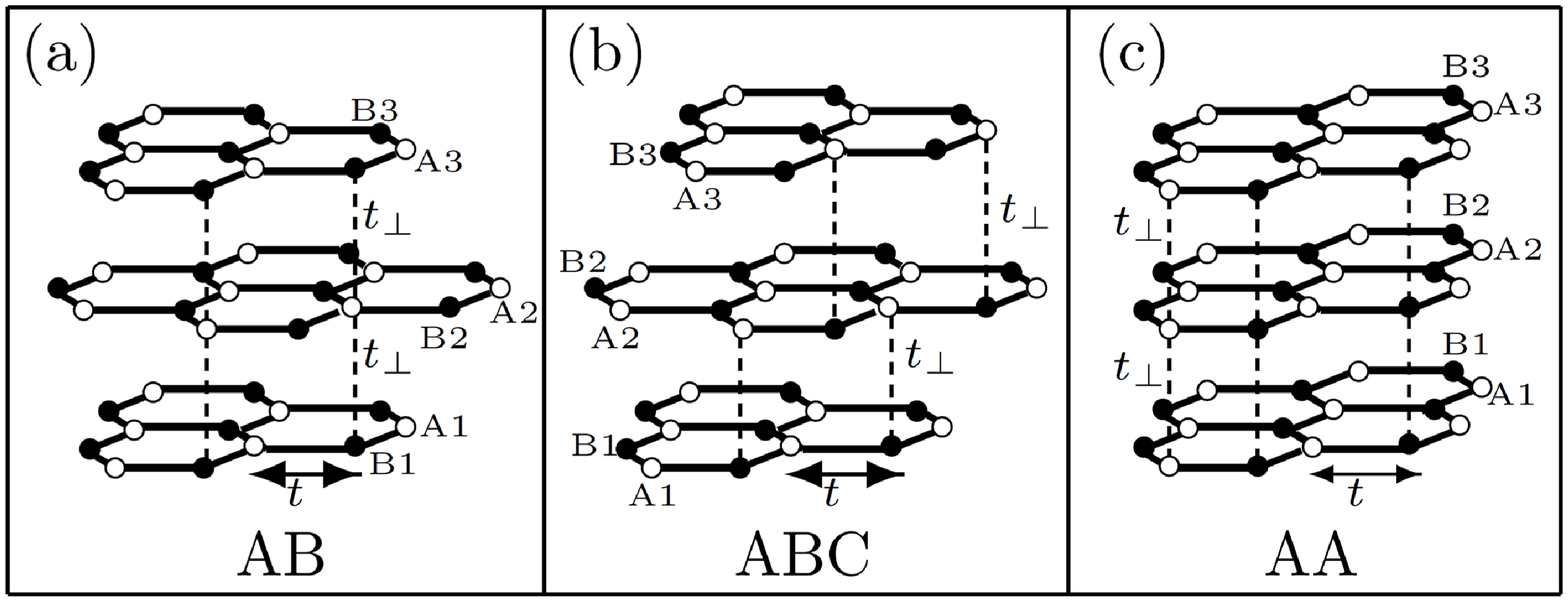}}}
\end{picture}
\hangcaption{Lattice structures of (a)AB, (b)ABC and
(c)AA-stacked trilayer graphenes, containing six sites in a unit
cell. White and black circles denote carbon atoms which belong to A and
B sublattices in each layer.} \label{fig2}
\end{figure}

---{\it Formalism}---
We consider three types of the stacking structures of
multilayer graphenes, AB, ABC and AA types.  Since a monolayer graphene
consists of two sublattices labeled by A and B, carbon atoms in
multilayer graphenes are specified by (A$_i$,B$_i$) meaning the A, B
sublattices in the $i$-th layer, respectively.  Figure~\ref{fig2} shows
the lattice structure of these systems with nearest neighbor intralayer
(interlayer) coupling $t$ ($t_{\perp}$).

Among these three types, we discuss AB stacked graphene in detail which is
the most common structure of graphite.  After taking the continuum
limit of tight-binding model, in a basis with atomic components for $N$
layers, $|{\rm A}_1\rangle,|{\rm B}_1\rangle,\cdots,|{\rm
A}_N\rangle,|{\rm B}_N\rangle$, the model Hamiltonian in the vicinity
of $K$ point (per spin and per valley) is
\begin{equation}
 \mathcal{H}
  =\left[
    \begin{array}{ccccc}
     H_0      & V & 0  & 0   & 0  \\
     V^\dagger  &      H_0 & V^\dagger    & 0  & 0   \\
     0      &  V & H_0  &  V  & 0  \\
     0    & 0   &V^\dagger  &      H_0    &\ddots   \\
     0     & 0   & 0      &\ddots    &\ddots
    \end{array}
\right],
\end{equation}
with
\begin{equation}
H_0
  =\left[
    \begin{array}{cc}
     0      & v\pi_-       \\
     v\pi_+ & 0   
    \end{array}
\right],
\quad\
V
  =\left[
    \begin{array}{cc}
     0      & 0       \\
     t_{\perp} & 0   
    \end{array}
\right],
\label{H_AB}
\end{equation}
where $\pi_{\pm}\equiv\pi_x\pm\i\pi_y$ with $\bm{\pi}\equiv
\bm{p}+e\bm{A}/c$ being the momentum operator in a magnetic field
$\nabla \times \bm{A}=(0,0,B)$. $v=(\sqrt{3}/2)\alpha t/\hbar $ is the
Fermi velocity with $\alpha $ being the lattice constant. We have
ignored long-range hopping terms except for $t$ and $t_{\perp}$ for
simplicity. Since the commutation relation between the momentum
operators in Eq.~(\ref{H_AB}) is $[\pi_\pm ,\pi_\mp ]=\mp 2eB\hbar /c$,
there are correspondences with the creation and annihilation operators
of the harmonic oscillator: $\pi_\pm \rightarrow \sqrt{2}\frac{\hbar
}{l}a^\dagger $ and $\pi_\mp \rightarrow \sqrt{2}\frac{\hbar }{l}a$ for
$eB\gtrless 0$, where $l\equiv \sqrt{c\hbar /|eB|}$.

In order to solve this model, we employ the matrix decompositions of
the Hamiltonian. It is already known that the Hamiltonians of AB and
AA-stacked $N$-layer graphenes can be block diagonalized, considering
Fourier modes of the wave function along the stacking direction
\cite{Koshino-mag,MacDonald,Nakamura-AA}. The same conclusion can also
be obtained by factorization of determinant of the Hamiltonians
\cite{Nakamura-multi}. According to these, the effective
Hamiltonian of AB-stacked $N$-layer graphenes can be divided into
isolated $[N/2]_G$ effective bilayer systems ($[x]_G$ is the integer
part of $x$), and one monolayer system is added when $N$ is odd.
Similarly, the effective Hamiltonian of an AA-stacked $N$-layer graphene
consists of $N$ isolated monolayer systems with different potential
energies.

Therefore, we can introduce a transformation matrix $U$ for the AB
stacked system which relates the wave function in the real space $|{\rm
A}_1\rangle,|{\rm B}_1\rangle,\cdots,|{\rm A}_N\rangle,|{\rm
B}_N\rangle$ and that in the Fourier modes of stacking direction
$|\phi_{N-1}^{\rm (A,even)}\rangle,|\phi_{N-1}^{\rm
(B,even)}\rangle,|\phi_{N-1}^{\rm (A,odd)}\rangle,|\phi_{N-1}^{\rm
(B,odd)}\rangle,|\phi_{N-3}^{\rm (A,even)}\rangle$,$\cdots$, $|\Psi
_\alpha \rangle ,|\Psi_\beta \rangle$. Here we have used the notation
defined in Refs.~\onlinecite{Koshino-mag} and~\onlinecite{Koshino-LL}.
Then the Hamiltonian ${\cal H}$ is transformed into a block diagonalized
form, {\renewcommand{\arraystretch}{2}
\begin{equation}
{\cal H'}=U^\dagger {\cal H}U=
 \left[	
  \begin{array}{c|c|c}
   {\cal H}^{\rm sub}(N-1) & & \\\hline
    &{\cal H}^{\rm sub}(N-3) & \\\hline
    & &\ddots                     \\
  \end{array}
 \right].
\end{equation}}
Here, ${\cal H}^{\rm sub}(m)$ is a Hamiltonian of a bilayer system with
an effective hopping
\begin{align}
&
{\cal H}^{\rm sub}(m)=\left[
        \begin{array}{cccc}
          0 &v\pi_- &0 &t_{\perp}\lambda_{m} \\
          v\pi_+ &0 &0 &0 \\
		  0 &0 &0 &v\pi_-  \\
		  t_{\perp}\lambda_{m} &0 &v\pi_+ &0 \\
		\end{array} \right],
\end{align}
with
\begin{displaymath}
 \lambda_{m}=2\sin\left(\frac{m\pi}{2(N+1)}\right),\quad
  m=N-1,N-3,\cdots >0.
\end{displaymath}
Using the above block diagonalized Hamiltonian, we easily obtain
eigenvalues of AB-stacked graphenes based on the known results for
monolayer and bilayer systems, replacing the interlayer hopping
$t_{\perp}$ by the effective one $t_{\perp}\lambda_{m}$,
\begin{align}
 \lefteqn{E^{\mu}_{m,n}
 =s_2 \frac{\sqrt{2}\hbar v}{l}
 \biggl\{ \frac{1}{2}\biggl( 2n+1+(\lambda_{m}r)^2} \nonumber\\
 &+s_1 \sqrt{(\lambda_{m}r)^4+2(2n+1)(\lambda_{m}r)^2+1}\biggl) 
 \biggl\}^\frac{1}{2}, \label{eigenbi}
\end{align}
where $r\equiv \frac{l}{\sqrt{2}\hbar v}t_{\perp}$ and $n$ denotes the
Landau levels.  The label $\mu\equiv(s_1,s_2)$ specifies the outer and
the inner bands ($s_1=\pm 1$), and positive and negative ($s_2=\pm 1$)
energies, respectively.  The eigenstates of AB-stacked graphenes in
basis of the real space $|{\rm A}_1\rangle,|{\rm
B}_1\rangle,\cdots,|{\rm A}_N\rangle,|{\rm B}_N\rangle$ are obtained
from those of the subsystems ${\cal H}^{\rm sub}(m)$ and the
transformation matrices $U$, written as 
\begin{equation}
 |\Psi_{n,\mu}\rangle =
  \left[ 
   \begin{array}{ccccc}
    f^{1}_{n,\mu}|n-1\rangle &
    f^{2}_{n,\mu}|n\rangle &
    f^{3}_{n,\mu}|n\rangle &
    f^{4}_{n,\mu}|n+1\rangle &
    \cdots
   \end{array} \right]^{T},
\end{equation}
where $f_{n,\mu}^{2k-1}$, $f_{n,\mu}^{2k}$ denote coefficients of the
wave function for the $k$-th layer, $|n\rangle$ is the number state of
$a, a^\dagger $, and $1\leq\mu\leq 2N$ is band indices of the multilayer
system.

\captionwidth=400mm

\begin{figure*}[t]
\begin{picture}(100,100)
\put(-100,20){
\put(-53,115){(a)}
\put(42,115){(b)}
\put(-56,-45){\scalebox{0.82}[0.9]{\includegraphics[width=10pc]{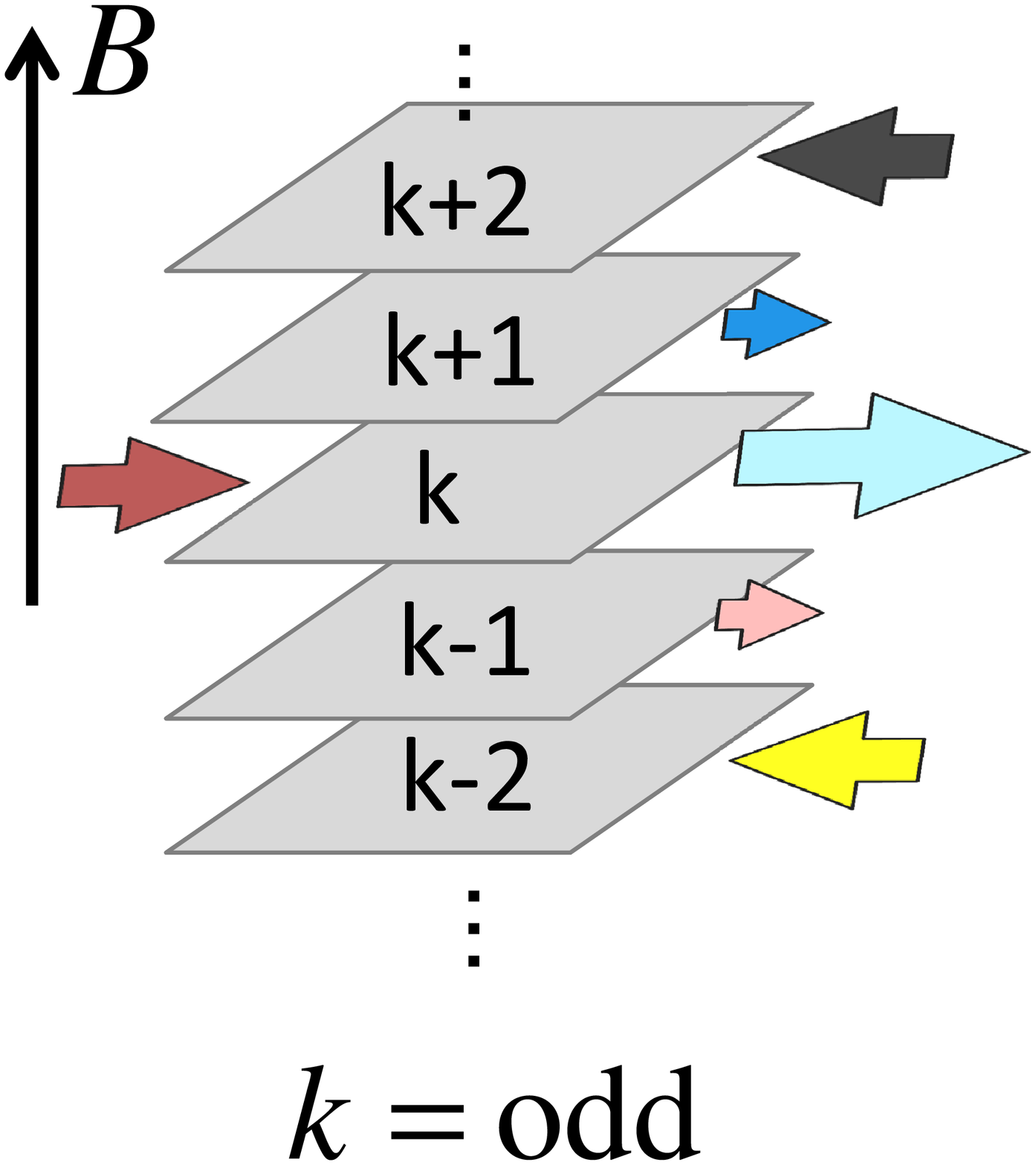}}}
\put(39,-45){\scalebox{0.82}[0.9]{\includegraphics[width=10pc]{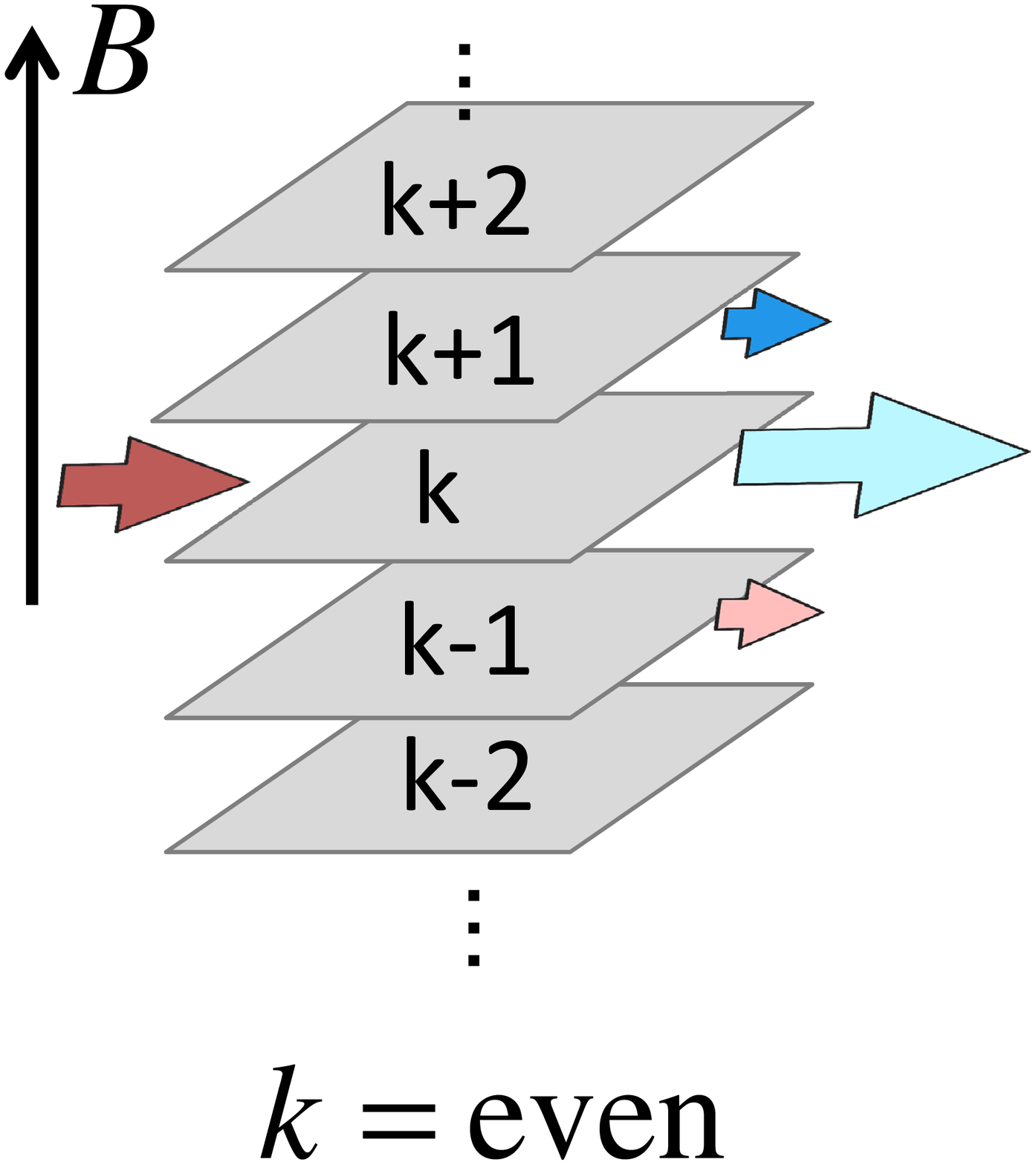}}}
}
\end{picture}
\begin{picture}(100,150)
\put(-35,142){(c)AB}
\put(69,142){(d)ABC}
\put(165,142){(e)AA}
\put(12,45){\line(5,2){8}}
\put(21,46){\scalebox{0.6}{$0.02t_\perp$}}
\put(11.5,38){\line(2,1){6}}
\put(18,40){\scalebox{0.6}{$0.01t_\perp$}}
\put(22,30.6){\line(1,1){3}}
\put(24,34.8){\scalebox{0.6}{$\mu \simeq 0$}}
\put(0,33){\rotatebox{90}{\scalebox{0.75}{$\sigma_{xx}^{1,4}$}}}
\put(19,17){\scalebox{0.7}{$B$(T)}}
\put(-62,93){\rotatebox{90}{\scalebox{0.75}{$\sigma_{xy}\Bigl(4e^2/h\Bigl)$}}}
\put(-62,26){\rotatebox{90}{\scalebox{0.75}{$\sigma_{xx}\Bigl(4e^2/\pi h\Bigl)$}}}
\put(102,-4){\scalebox{1.2}{$\mu /\sqrt{B'}$}}
\put(-62,75){\includegraphics[width=9.8pc]{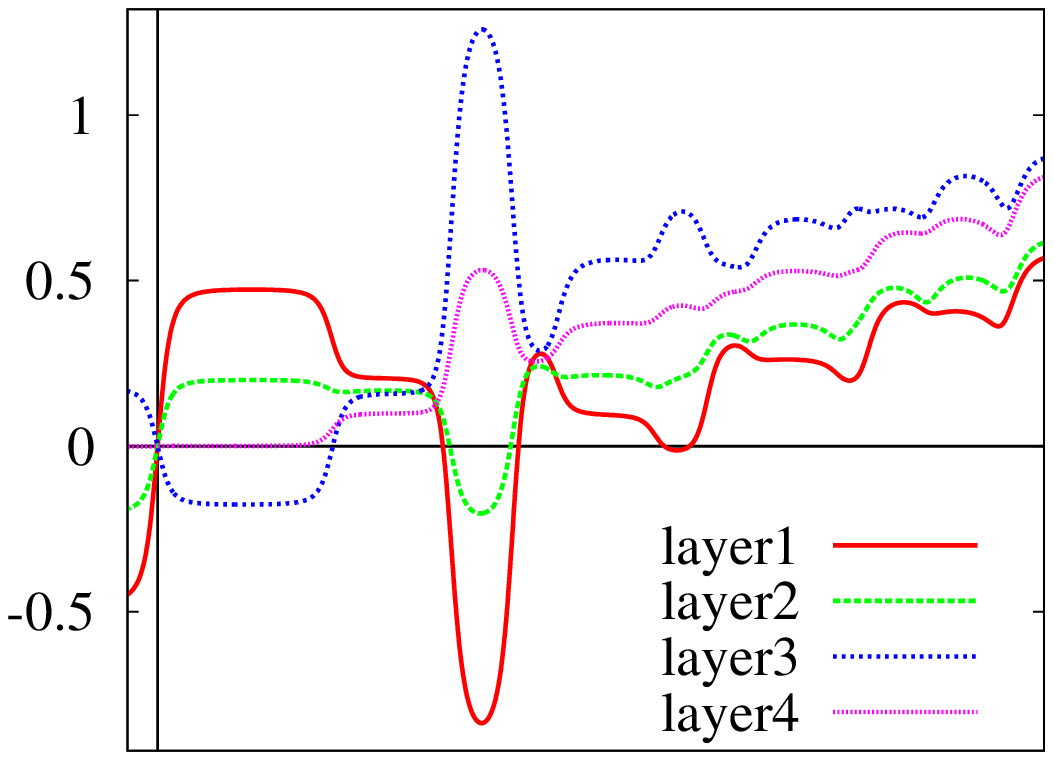}}
\put(40,75){\includegraphics[width=9.8pc]{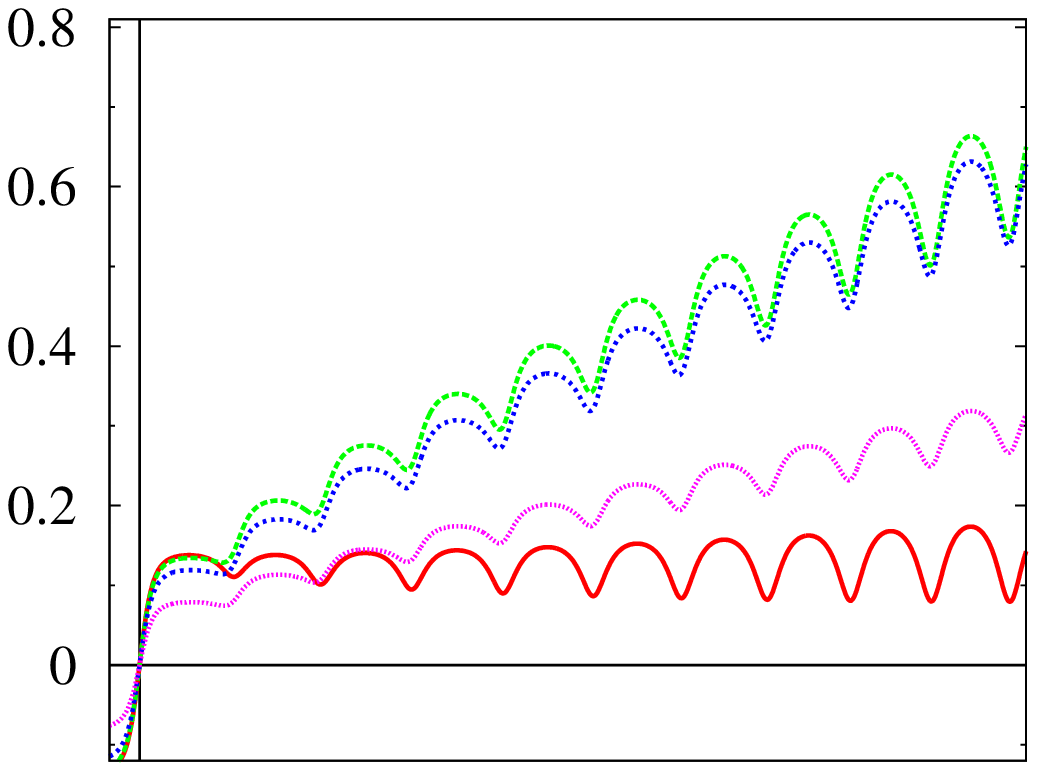}}
\put(143,75){\includegraphics[width=9.8pc]{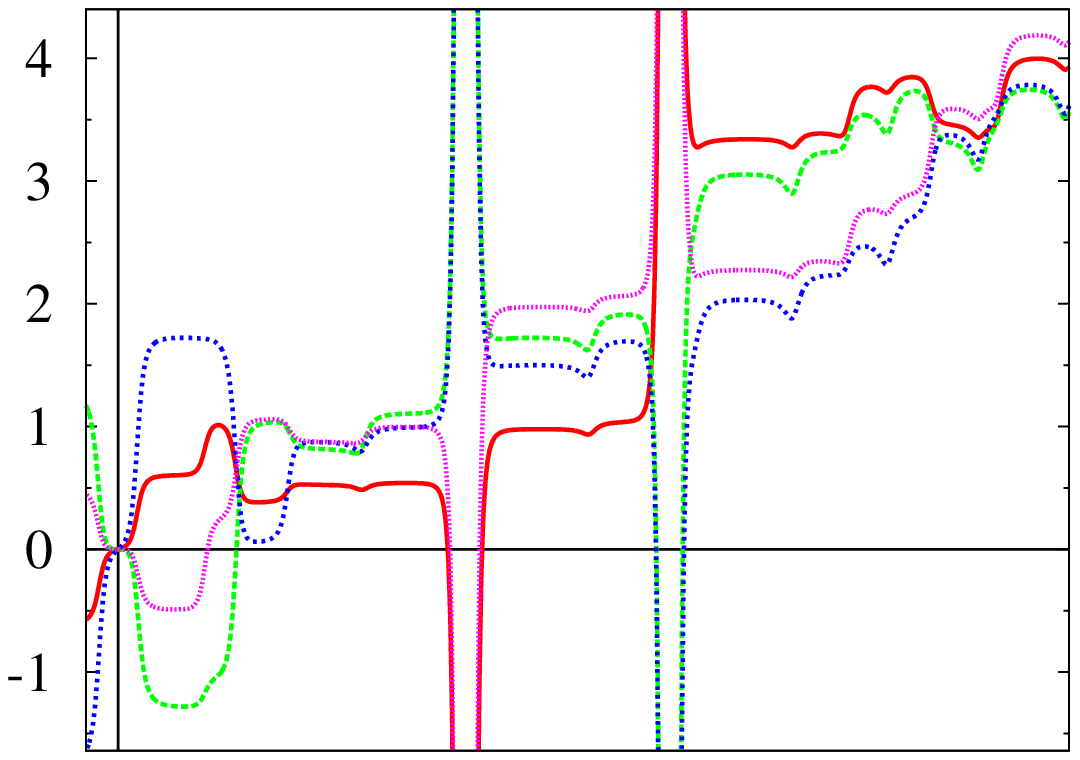}}
\put(-62,5.3){\includegraphics[width=9.8pc]{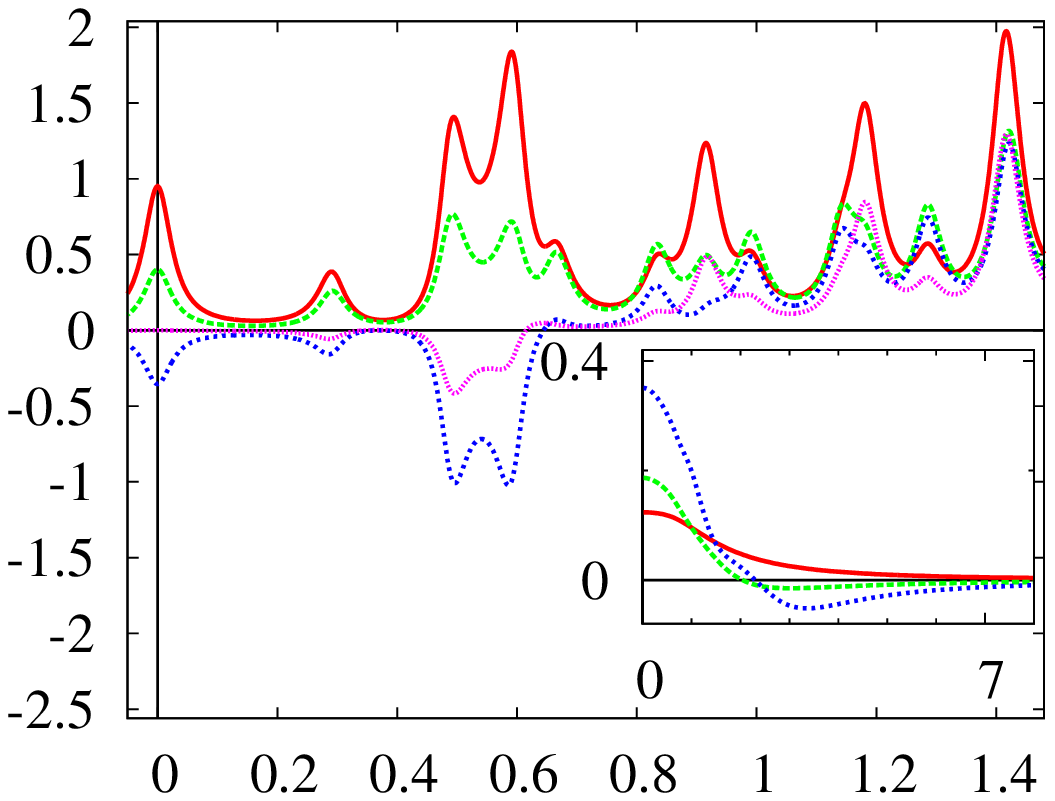}}
\put(40,5.3){\includegraphics[width=9.8pc]{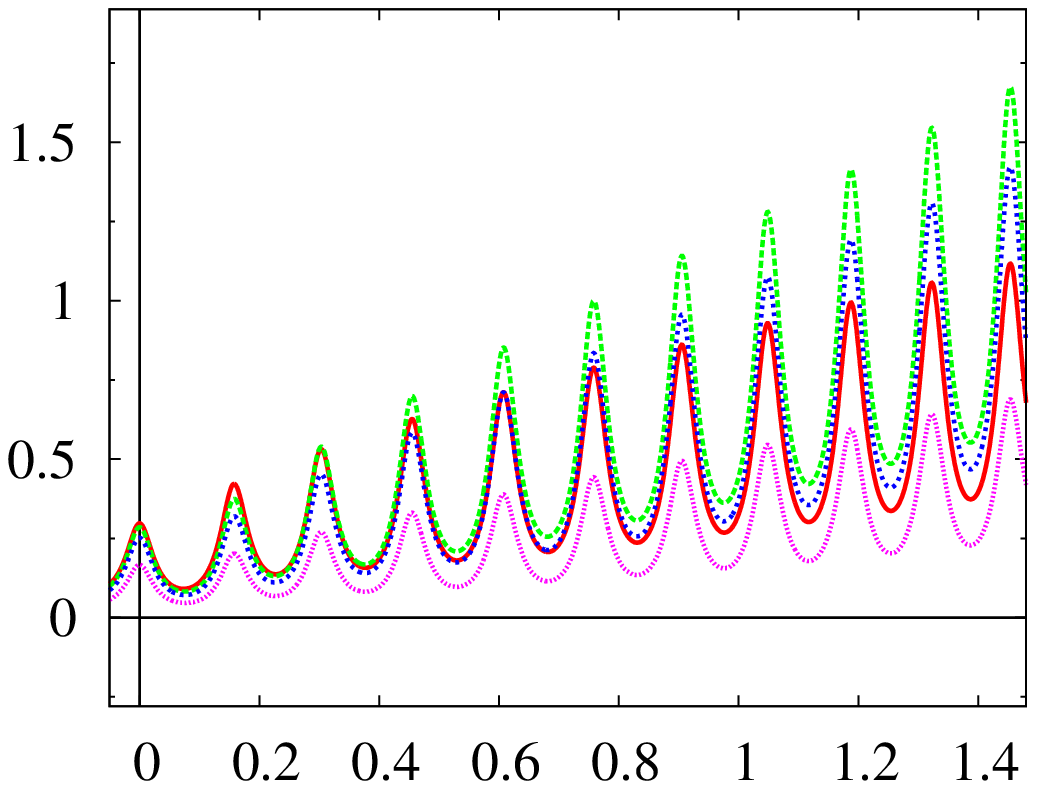}}
\put(143,5.3){\includegraphics[width=9.8pc]{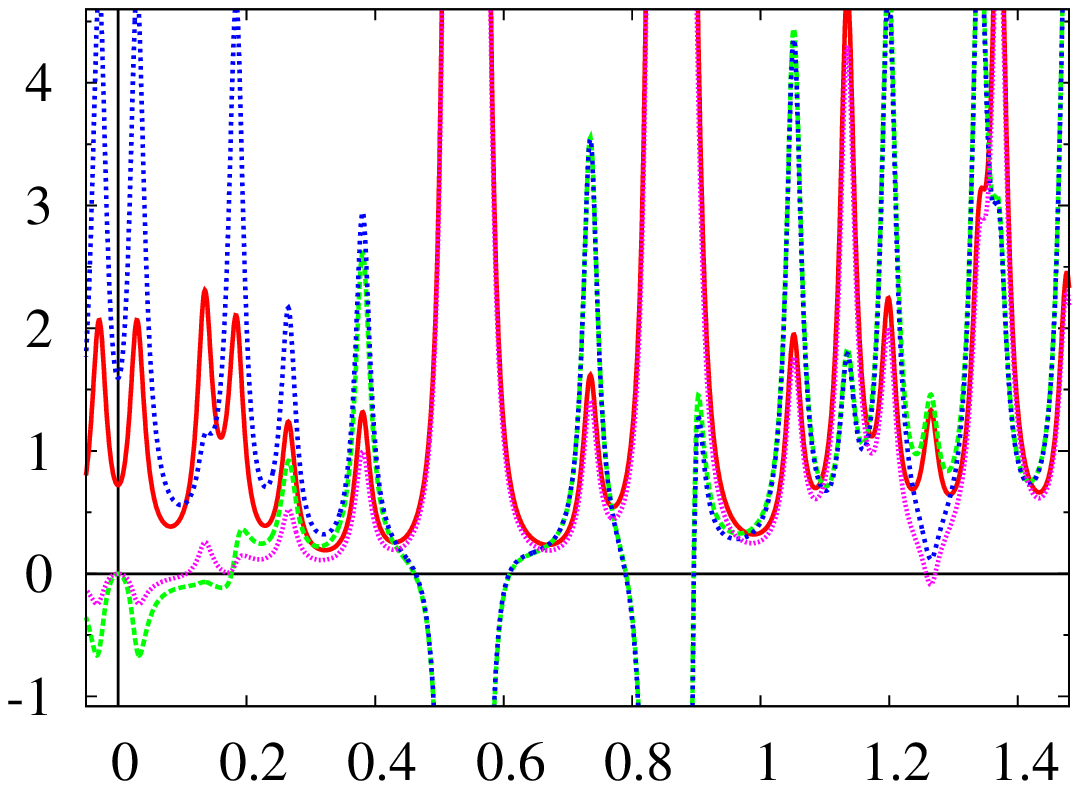}}
\end{picture}
\caption{Schematic illustration of intensity of the layer-resolved
conductivity for AB-stacked multilayer graphene applying bias to the
$k$-th layer where $k$ is (a) odd and (b) even layers counting along the
direction of the magnetic field. The layer-resolved Hall
($\sigma_{xy}^{kl}$) and longitudinal ($\sigma_{xx}^{kl}$)
conductivities of 4-layer graphene in magnetic fields ($B=14$T) for
(c)AB, (d)ABC and (e)AA stacking.  The bias is applied to the 1st layer.
An inset in (c) shows $\sigma_{xx}^{14}$ as functions of the magnetic
field.}\label{fig.3}
\end{figure*}

The conductivity is given by the Kubo formula as
\begin{equation}
 {\rm Re }\hspace{1mm}\sigma_{ij}(\Omega )
  =-\frac{\Im\Pi_{ij}(\Omega)}{\hbar\Omega},
\end{equation}
where $\Pi_{ij}$ with $\{ i,j\} \in \{x,y\}$ and $\Pi_{ij}$ is the Fourier
transform of the current-current correlation function obtained after 
analytic continuation of the Matsubara form.  The general expression of
the conductivity of the multilayer graphenes is obtained by extending the
result for the bilayer graphene \cite{Nakamura-bi} as
\begin{align}
 &\sgn (eB)\sigma_{xy}(\Omega)+\i\sigma_{xx}(\Omega)
 =-\frac{4e^2v^2}{h l^2\Omega}
 \sum_{n}\sum_{\mu,\nu}\label{sigma_xyxx}\\
&
 \left\{
 X(E^{\mu}_{n+1},E^{\nu}_{n};\Omega)-
 X(E^{\mu}_{n+1},E^{\nu}_{n};-\Omega)
 \right\}
 \biggl(\sum\limits_{k=1}^{N}f_{n+1,\mu }^{2k-1}f_{n,\nu }^{2k}
 \biggl)^2,\nonumber
\end{align}
with
\begin{align}
X(A,B;\Omega)\equiv&\sum\limits_n[(\i\tilde{\omega}_n-A/\hbar)^{-1}
 (\i\tilde{\omega}_{n+m}-B/\hbar)^{-1} |_{\i\nu_m\rightarrow
 \Omega}\nonumber\\ &\qquad\quad -(\Omega \rightarrow -\Omega)](\Omega
 \beta \hbar)^{-1},
\end{align}
where $\tilde{\omega}_n$ is Matsubara frequency of fermion including the
chemical potential and effect of impurity scattering $\Gamma $ as
$\i\tilde{\omega}_n+[\mu +\i\,\sgn (\omega_n)\Gamma]/\hbar$.  $\nu_m$ is
the bosonic Matsubara frequency and $\omega_{n+m}\equiv \omega_n+\nu_m$.

In the above formalism, the electric current operators are defined by
$J_i=-\frac{\delta \mathcal{H}}{\delta A_i}$ with $A_i$ the vector
potential in the direction $i=x,y$.  Now, we introduce current operators
for particular layers to calculate the conductivity between two distinct
layers as $J_{i}^k=-\frac{\delta \mathcal{H}}{\delta A_{i}^k}$ where $k$
denotes layer number ($k=1,2,\cdots, N$), and $A_{i}^k$ is the vector
potential acting solely in layer $k$ in direction $i$.  The current
operator of the $k$-th layer has only two matrix elements,
$(J^k_i)_{2k-1,2k}$ and $(J^k_i)_{2k,2k-1}$, and all other elements are
vanishing. The general expression for the layer-resolved conductivity
between the $k$-th and $l$-th layers, $\sigma_{ij}^{kl}$ is given by
Eq.~(\ref{sigma_xyxx}) with the following replacement,
\begin{align}
 \biggl(\sum\limits_{k=1}^{N}f_{n+1,\mu }^{2k-1}f_{n,\nu }^{2k}
 \biggl)^2
\to
 f_{n+1,\mu}^{2k-1}f_{n,\nu}^{2k}
 f_{n+1,\mu}^{2l-1}f_{n,\nu}^{2l}.
\end{align}
In the present model which includes only the nearest interlayer
couplings as interlayer matrix elements, the layer current operator,
$J_{i}^k$ does not have any momentum dependence.  However, this
situation can change if we introduce other matrix elements, such as
tilted interlayer hoppings.  In such cases, the layer current operators
should be redefined appropriately so that they become Hermitian and
satisfy the relation $J_{i}=\sum_{k=1}^N J_{i}^k$.

---{\it Numerical results}---
Based on the above discussions, we calculate the layer resolved Hall
conductivity $\sigma_{xy}^{kl}$ of AB-stacked multilayer graphenes, as
functions of the Fermi energy, for strong magnetic field
$\hbar\omega_{\rm c}\gg\Gamma$ where $\omega_{\rm c}\propto \sqrt{B}$ is
the cyclotron frequency.  Our findings are summarized as: i)
$\sigma_{xy}^{kl}$ is no longer quantized as integer times $e^2/h$.  ii)
around the zero energy $\mu\sim 0$, $\sigma_{xy}^{kl}$ becomes almost
zero when the source and the drain are separated by more than three
layers $|k-l|\geq 3$ and the biased layer $k$ is odd, and two layers
$|k-l|\geq 2$ and $k$ is even.  iii) the layer resolved Hall response
can be negative ($\sigma_{xy}^{kl}<0$) around the zero energy for
$|k-l|=2$ when the biased layer $k$ is odd.  Layers indices are growing
in the direction of the magnetic field.  These features are summarized
in Fig.~\ref{fig.3}(a),(b). Similar behavior can also be seen in the
longitudinal conductivity $\sigma_{xx}^{kl}$ which behaves like
derivatives of $\sigma_{xy}^{kl}$ by $\mu$.

In Fig.~\ref{fig.3} we show numerical results of $\sigma_{xy}^{kl}$ for
$N=4$ as functions of the Fermi energy $\mu$, where the bias is applied
to the first layer, and $B'\equiv2v^2eB/c$.  We have assumed $B=14$T,
$T=0$K, $\Gamma =0.01t_{\perp}$, $t=3.16$eV, $t_{\perp}=0.39$eV, and
taken DC limit $\Omega \rightarrow 0$.  We can see the above three
features for AB staking in Fig.~\ref{fig.3}(c). Further,
$\sigma_{xx}^{14}$ is finite without magnetic field, meaning that
property ii) is broken when magnetic field is turned off, as shown in
the inset of Fig.~\ref{fig.3}(c). The conductivity vanishes when
$B^\prime \sim \Gamma$.

We also calculate layer resolved conductivities for other stacking
structures.  For ABC stacking, although the matrix decomposition
technique is no longer available, we can diagonalize the Hamiltonian
numerically to obtain the layer resolved conductivities.  In this case, as
shown in Fig.~\ref{fig.3}(d), the electric currents are induced in every
layer.  Moreover, the conductivities for ABC-stacking do not exhibit
negative values even in the vicinity of zero energy.

For AA stacking, the matrix decomposition technique similar to AB
stacking can also be used \cite{Nakamura-AA,MacDonald}. We calculate the
conductivities by setting the value of interlayer hopping as
$t_{\perp}^{\rm AA}=t_{\perp}^{\rm AB, ABC}/2$.  In this system,
negative conductivity also appears as shown in Fig.~\ref{fig.3}(e). In
this case, it is difficult to summarize the features of layer resolved
conductivities in terms of simple rules, because the effective
Hamiltonian consists of $N$ monolayer systems with different Fermi
energies, so that the number of Landau levels near the zero energy
depends strongly on the strength of the magnetic field. In contrast to
this, the Landau level structure is an intrinsic property for AB and
ABC-stacked systems.


---{\it Analytical results for AB stacking}---
In order to understand the above results for AB stacking in more detail,
we calculate the first quantum Hall step of $\sigma_{xy}^{kl}$
analytically for $\Gamma=0$. We consider the following four cases: (a/b)
$(k,l)$=(odd, odd) for $N=$even/odd, respectively, (c) $(k,l)$=(odd,
even), (d) $(k,l)$=(even, even).  Since layer indexes grow in the
direction of the field, such classification is unique.  We obtain the
following results, defining $\tilde{\sigma}^{k,l}_{xy}\equiv
\sgn(eB)\frac{h}{4e^2}\sigma^{k,l}_{xy}$,
\begin{subequations} 
\begin{align}
 \tilde{\sigma}_{xy}^{k,l({\rm a})}=&
 \left[
 \frac{r^2}{4}(\delta_{l',k'\pm 1}
 -\delta_{k',1}\delta_{l',1})+\frac{1+r^2}{2}\delta_{k',l'}
 \right] \nonumber\\
 &\times \sum_m 
 \frac{U_{4k'-2,m}U_{4l'-2,m}}{1+(\lambda_mr)^2}
\\
 \tilde{\sigma}_{xy}^{k,l({\rm b})}
 =&\bigg[ \frac{r^2}{4}(\delta_{l',k'\pm 1}
 -\delta_{k',1}\delta_{l',1}
 -\delta_{k',\frac{N+1}{2}}\delta_{l',\frac{N+1}{2}})\nonumber\\
 &+\frac{1+r^2}{2}\delta_{k',l'}\bigg]\nonumber\\
& \times \bigg[ \sum_m 
 \frac{U_{4k'-2,m}U_{4l'-2,m}}{1+(\lambda_mr)^2}
 +\frac{2(-1)^{k'+l'}}{N+1}\bigg],\\
 \tilde{\sigma}_{xy}^{k,l({\rm c})}
 =&\frac{r^2}{4}
 (\delta_{l',k'}+\delta_{l',k'-1})
 \sum_m \frac{U_{4k'-2,m}U_{4l',m}\lambda_m}{1+(\lambda_mr)^2}, \\
 \tilde{\sigma}_{xy}^{k,l({\rm d})}
 =&
 \delta_{k',l'}
 \sum_m 
 U_{4k',m}U_{4l',m}\frac{2+(\lambda_mr)^2}{4(1+(\lambda_mr)^2)}.
\end{align}
\end{subequations}
where $k'\equiv\left[\frac{k+1}{2}\right]_G$ and
$l'\equiv\left[\frac{l+1}{2}\right]_G$.  From these results, it is
apparent that interlayer conductivity is finite only within nearest or
second nearest layers (property ii)).  For the simplest example, the
values of $\tilde{\sigma}^{k,l}_{xy}$ for three layer system obtained
from the above formalism are
$\tilde{\sigma}^{1,1}_{xy}=(2+3r^2+r^4)/(4+8r^2)$,
$\tilde{\sigma}^{1,2}_{xy}=r^2/(4+8r^2)$,
$\tilde{\sigma}^{2,2}_{xy}=(2+2r^2)/(4+8r^2)$,
$\tilde{\sigma}^{1,3}_{xy}=-r^4/(4+8r^2)$, with
$\tilde{\sigma}^{2,3}_{xy}=\tilde{\sigma}^{1,2}_{xy}$ and
$\tilde{\sigma}^{3,3}_{xy}=\tilde{\sigma}^{1,1}_{xy}$.  In this case,
$\sigma^{1,3}_{xy}$ becomes negative. By summing up all
contributions, we obtain $\sum_{k,l}\tilde{\sigma}^{k,l}_{xy}=3/2$ which
consists of contributions from effective monolayer ($1/2$) and bilayer
($1$).  For other $N$ cases, these analytical results coincide with the
general results summarized in Fig.~\ref{fig.3}(a) and (b).


---{\it Conclusion and discussion}---
To conclude, we have studied the interlayer electronic transport
properties of multilayer graphenes in a magnetic field for variety of
stacking orders.  The behaviour of the layer-resolved
conductivity depends strongly on the stacking
structure. For AB stacking, various interesting
properties, such as negative response, and suppression of the Hall
conductivity, are identified.  The breakdown of the quantization of the Hall
conductivity indicates that interlayer conductivity has different
features from those of the total response, and is not protected by topology.

Finally, we discuss the possibility of the experimental observation of the
layer-resolved conductivities.  Though it would be rather challenging to
connect leads to particular layers, it is certainly easier to measure the
layer-resolved conductivity between the top and the bottom layers. 
The fact that the longitudinal conductivity greatly changes
according to the magnetic field (for example $\sigma_{xx}^{1,4}$ as shown
in the inset of Fig.~\ref{fig.3}(c)) means that multilayer systems may
be applied as switching device. We think that our work provides a
comprehensive understanding of transport properties of multilayer
graphene.

\smallskip

---{\it Acknowledgments}---
We thank D.-H. Lee, M.~Oshikawa, H.~Shimada, and Y.~Tada for
discussions. M. N. acknowledges support from Global Center of Excellence
Program ``Nanoscience and Quantum Physics'' of the Tokyo Institute of
Technology and Grants-in-Aid No.23540362 by MEXT.  B.D. was supported by
the Hungarian Scientific Research Fund No. K72613, K73361, CNK80991, New
Sz\'echenyi Plan Nr.  T\'{A}MOP-4.2.1/B-09/1/KMR-2010-0002 and by the
Bolyai program of the Hungarian Academy of Sciences is acknowledged.



\end{document}